\documentclass{PoS}

\title{Novel QCD Phenomena at JLab}

\ShortTitle{Novel QCD Phenomena at JLab}

\author{\speaker{Stanley J. Brodsky}\\
        SLAC National Accelerator Laboratory\\
        Stanford University\\
        E-mail: \email{sjbth@slac.stanford.edu}}

\abstract {The $12~$GeV electron beam energy at Jefferson Laboratory ($\sqrt s \simeq 4.8 ~ $ GeV ) provides ideal electroproduction kinematics for many novel tests of QCD in both the perturbative and nonperturbative domains.  These include tests of the quark flavor dependence of the nuclear structure functions;  measurements of the QCD running coupling at soft scales;  measurements of the diffractive deep inelastic structure function;  measurements of  exclusive contributions to the $T-$ odd Sivers function; 
the identification of ``odderon"  contributions; tests of the spectroscopic and dynamic features of light-front holography as well as ``meson-nucleon supersymmetry";  the production of open and hidden charm states in the heavy-quark threshold domain;  and the production of exotic hadronic states such as pentaquarks, tetraquarks and even octoquarks containing charm quarks.  One can also study fundamental features of QCD at JLab$12$ such as the ``hidden color" of nuclear wavefunctions, the ``color transparency" of hard exclusive processes, and the ``intrinsic strangeness and charm" content  of the proton wavefunction.  In addition, current on-going experimental searches for ``dark photons" at JLab can also be used to produce and detect a long-sought exotic atom of QED: ``true muonium"  $[\mu^+ \mu^-].$   I will also discuss evidence that the antishadowing of nuclear  structure functions is non-universal;  i.e.,  flavor dependent.  I will also present arguments why shadowing and antishadowing phenomena may be incompatible with the momentum and other sum rules for the nuclear parton distribution functions.  I will also briefly review new insights into the hadron mass scale, the hadron mass spectrum, the functional form of the QCD coupling in the nonperturbative domain predicted by light-front holography, and how superconformal algebra leads to remarkable supersymmetric relations between mesons and baryons.}

\FullConference{QCD Evolution 2015 -QCDEV2015-\\
		26-30 May 2015\\
		Jefferson Lab (JLAB), Newport News Virginia, USA}

\begin{document}

\section{Introduction}
The new $12~$ GeV electron beam at Jefferson Laboratory~\cite{Smith:2009yk} provides ideal kinematics for testing many aspects quantum chromodynamics (QCD), the fundamental theory of hadron physics.  Electroproduction experiments in the JLab$12$ energy range $W = \sqrt s \simeq 4.8 $ GeV are above the threshold for producing open and hidden charm states, including the production on protons and nuclei of exotic hadronic states such as pentaquarks, tetraquarks, and even 
``octoquarks" containing charm quarks~\cite{Brodsky:2015wza,Brodsky:1987xw,Bashkanov:2013cla}.

Jlab12 can also provide ideal electroproduction kinematics for many novel tests of QCD in both the perturbative and nonperturbative domains~\cite{Brodsky:2012zzb}.  These include tests of the flavor dependence of the antishadowing of nuclear structure functions~\cite{Brodsky:1989qz,Brodsky:2004qa,Brodsky:2005ww}; measurements of the QCD running coupling (effective charge) in the nonperturbative domains~\cite{Brodsky:2010ur};   measurements of  diffractive deep inelastic  scattering~\cite{Brodsky:2002ue};  measurements of  exclusive contributions to the $T-$ odd Sivers function~\cite{Brodsky:2002cx}; 
 the identification of ``odderon"  contributions in $\pi^0$ photoproduction and odderon-Pomeron interference  from the $s$ vs $\bar s$ asymmetry in $\gamma^* p \to s \bar s X$ ~\cite{Brodsky:1999mz}; verification of the $``J=0"$ energy-independent real part of Compton amplitude in $\gamma^* p \to \gamma p^\prime$ via interference with Bethe-Heitler amplitude~\cite{Brodsky:2009bp,Brodsky:1973hm};
tests of the spectroscopic and dynamic features of light-front holography~\cite{Brodsky:2014yha}, 
as well as ``meson-nucleon supersymmetry"~\cite{Dosch:2015nwa}.    One can also study fundamental features of QCD at JLab$12$ such as the `
``hidden color" of nuclear wavefunctions~\cite{Brodsky:1983vf}, the ``color transparency" of hard exclusive processes~\cite{Brodsky:1988xz} and the ``intrinsic strangeness and charm" content of the proton wavefunction~\cite{Brodsky:1984nx,Brodsky:1980pb,Brodsky:2015uwa}.  In addition, current on-going experimental searches for ``dark photons" at JLab can also be used to produce and detect a long-sought exotic atom of QED: ``true muonium" $[\mu^+ \mu^-]$~\cite{Brodsky:2009gx,Banburski:2012tk}.

The JLab fixed target facility is an important compliment to a high intensity electron-ion collider. The  proposed LHeC electron-proton collider~\cite{AbelleiraFernandez:2012cc} at CERN will extend electroproduction studies to the multi-TeV domain where the collisions of top quarks, and electroweak bosons can be studied.  One studies the same boost-invariant light-front wavefunctions of the rest-frame proton and the ultrarelativistic proton at the LHeC despite very different kinematics --  the electron scatters on the quarks of the proton at fixed light-front time  $\tau=t+z/c$~\cite{Dirac:1949cp,Brodsky:1997de},  where Lorentz boosts are kinematical,  not   at a fixed ``instant" time $t$, in which case boosts are dynamical and can even change particle number.

I will also discuss  in this report the indications that the antishadowing of nuclear structure functions is nonuniversal, i.e., 
flavor dependent~\cite{Schienbein:2007fs,Brodsky:1989qz,Brodsky:2004qa}. I will also present arguments why shadowing and antishadowing phenomena are  incompatible with momentum and other sum rules for nuclear PDFs.  I will also briefly review new insights into the hadron mass scale, the hadron mass spectrum, and the effective QCD effective charge in the nonperturbative domain which have been obtained from light-front holography~\cite{deTeramond:2008ht,Brodsky:2014yha}. Finally, I will show how 
superconformal  quantum mechanics~\cite{Fubini:1984hf}  predicts  supersymmetric relations between mesons and baryons in QCD~\cite{Dosch:2015nwa}. 

\section{Is Antishadowing Flavor-Dependent? }

\begin{figure}
 \begin{center}
\includegraphics[height=9.5cm,width=15cm]{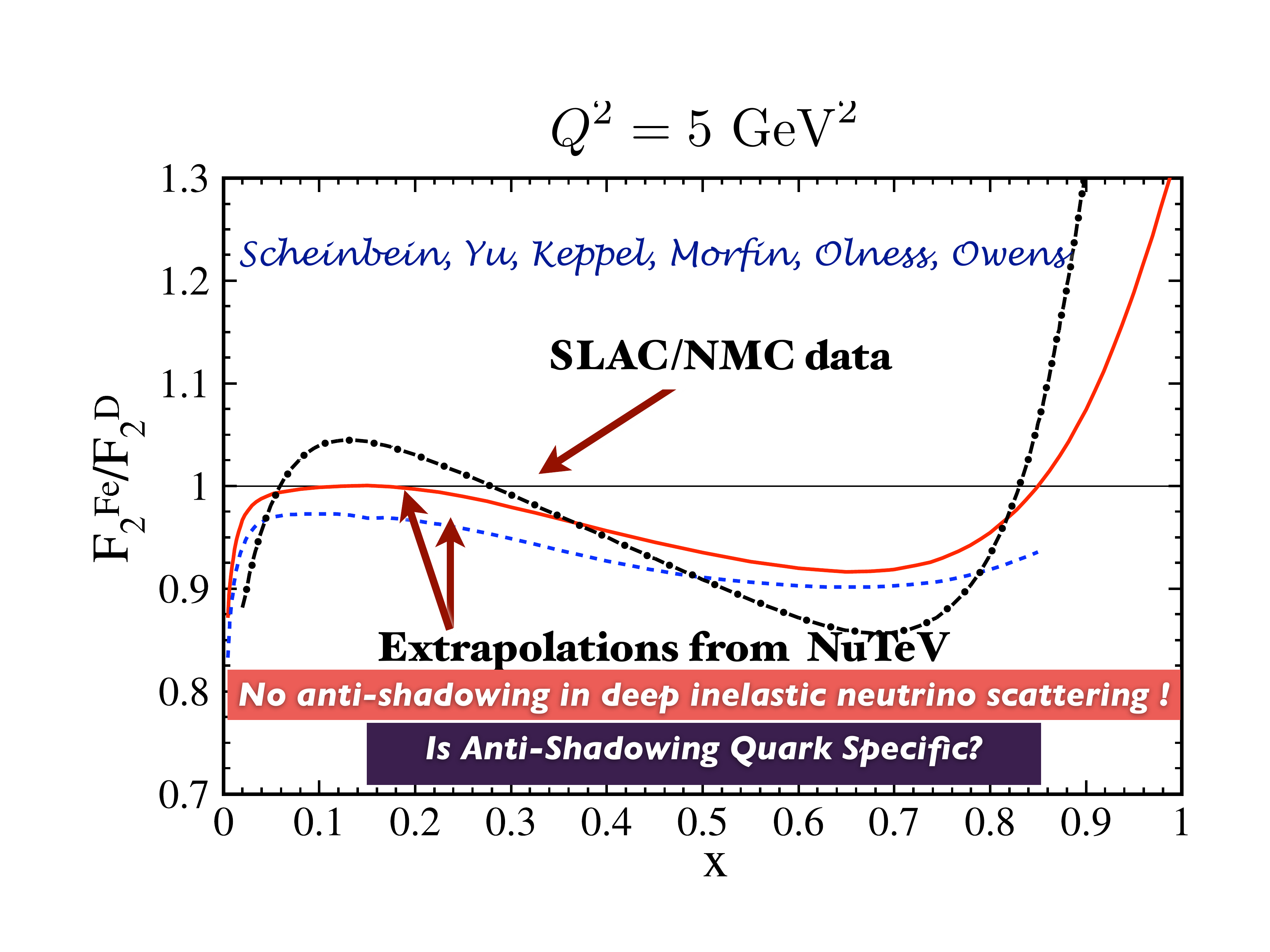}
\end{center}
\caption{Comparison of  nuclear structure functions measured in charge and neutral current deep inelastic lepton scattering. 
The NuTeV charged current measurement $\mu A \to \nu X $ scattering does not appear to show antishadowing. 
The compilation is from Scheinbein {\it et al}.~\cite{Schienbein:2007fs}.}
\label{FigsJlabProcFig1.pdf}
\end{figure}

The ratio of the iron to deuteron nuclear structure functions $F^{Fe}_2(x,Q^2)/ F^{D}_2(x,Q^2)$  measured 
in charge-current ($W^*$ exchange) and neutral-current  ($\gamma^* $ exchange) deep inelastic lepton scattering 
is illustrated in fig.~\ref{FigsJlabProcFig1.pdf}.
Although the SLAC and NMC deep inelastic lepton scattering data shows both shadowing and antishadowing, the NuTeV measurement of the charged current reaction $\mu A \to \nu X $ does not appear to show antishadowing. This important observation by Scheinbein {\it et al.},~\cite{Schienbein:2007fs}  is in direct contradiction to the usual assumption that the nuclear structure function measures the square of the nuclear light-front wavefunction and is thus  independent of the probe.   It is also  contradicts the usual assumption that the nuclear modifications of shadowing and antishadowing 
must  balance each other~\cite{Nikolaev:1975vy} in order to satisfy the momentum sum rule.  

In fact, these assumptions are wrong in the Gribov-Glauber~\cite{Glauber:1955qq} description of shadowing and shadowing.   
In the Gribov-Glauber theory~\cite{Stodolsky:1966am,Brodsky:1969iz,Brodsky:1989qz}, the shadowing of the nuclear structure function 
measured in deep inelastic lepton-nucleus colisions is due to the {\it destructive } interference  of two-step and one-step processes. 
In the two-step process, the incoming current first produces diffractive deep inelastic scattering (DDIS) on a nucleon $N_a$ located on the front surface of the nucleus: 
$\gamma^*  N_a \to V N^\prime_a$, where $N_a$ scatters intact  and stays in the nuclear ground state.  
The vector meson  system $V$ then interacts inelastically on a second nucleon $N_b$ in the nuclear interior $V + N_b \to X$.    
This two-step process interferes coherently and destructively with the usual one-step DIS amplitude $\gamma^* N_b \to X$ where in this case $N_a$ is a spectator. 
The interference is destructive since the diffractive DIS  amplitude due to Pomeron exchange has phase $i$, and the Glauber cut gives another factor of $i$.  
Thus the interior nucleon $N_b$ sees two beams:  the primary virtual photon beam and the secondary vector hadronic system $V$.   In effect, the interior nucleon $N_b$ is `shadowed' by the front-surface nucleon $N_a$.   One can account for the magnitude and shape of the observed nuclear shadowing from this mechanism~\cite{Brodsky:1989qz,Brodsky:2004qa,Brodsky:2005ww};.

In contrast, antishadowing  can be understood as arising from the {\it constructive interference } of the two-step and one-step processes~\cite{Brodsky:1989qz,Brodsky:2004qa}.
In this case  the two-step process  involves  the contribution to diffractive DIS $\gamma^*  N_a \to V N^\prime_a$  from isospin $I=1$ Reggeon exchange in the $t$ channel. This  Reggeon with $\alpha_R \simeq 1/2$  accounts for the Kuti-Weisskopf behavior 
$F^p_2(x,Q^2) - F^n_2(x,Q^2) \propto  x_{bj}^{0.5}$ at small $x_{bj}$~\cite{Kuti:1971ph}. 
The phase of the Reggeon amplitude  - its ``signature factor" -- is determined by charge conjugation and crossing.  The result is that in this case, the two-step and one-step amplitudes interfere {\it constructively}.  One can then account for the shape and magnitude of the  antishadowing enhancement of the nuclear structure function measured in $e A \to e^\prime X$ in the $0.1 < x_{bj} <   0.2$ domain.  However, the physics of antishadowing is flavor dependent  since each quark and anti-quark has its own specific coupling to the $I=1$ Reggeon.  Thus antishadowing of nuclear structure function, unlike Pomeron-dominated shadowing, is predicted to be flavor dependent~\cite{Brodsky:1989qz,Brodsky:2004qa,Brodsky:2005ww}.    Since the weights of quark flavors are different, the antishadowing of the nuclear structure function measured in the charged current DIS NuTeV reaction can differ from the antishadowing measured in the SLAC and NMC neutral current reaction.
This novel physics could be tested at JLab-12 by tagging the flavor of the struck quark in the $e q \to e^\prime q^\prime$ hard subprocess.

\section{Is the Momentum Sum Rule Valid for Nuclear Structure Functions? }

Sum rules for DIS  processes are analyzed using the operator product expansion of the forward virtual Compton amplitude, assuming it reduces in the limit $Q^2 \to \infty$ to matrix elements of local operators such as the energy-momentum tensor.  The moments of the structure function and other distributions can then be evaluated as overlaps of the target hadron's light-front wavefunction (LFWF), the hadronic eigensolution of the LF Hamiltonian,  as in the Drell-Yan-West formulae for hadronic form factors~\cite{Brodsky:1980zm,Liuti:2013cna,Mondal:2015uha,Lorce:2011dv}.
The phases of the resulting DIS amplitude and OPE matrix elements reflect the real phase of the stable target hadron's wavefunction.
This approximation defines the ``static"  contribution~\cite{Brodsky:2008xe,Brodsky:2009dv} to the measured parton distribution functions (PDF), transverse momentum distributions, etc.  The resulting momentum, spin and other sum rules reflect the properties of the hadron's light-front wavefunction.

However, final-state interactions which occur {\it after}  the lepton-quark scattering, give non-trivial contributions to deep inelastic scattering processes at leading twist and survive at high $Q^2$ and high $W^2 = (q+p)^2.$  For example, the pseudo-$T$-odd Sivers effect~\cite{Brodsky:2002cx} is directly sensitive to the rescattering of the struck quark. 
Similarly, diffractive deep inelastic scattering involves the exchange of a gluon after the quark has been struck by the lepton~\cite{Brodsky:2002ue}. 
These ``lensing" corrections survive when both $W^2$ and $Q^2$ are large since the vector gluon couplings grow with energy. Part of the phase can be associated with a Wilson line as an augmented LFWF~\cite{Brodsky:2010vs} which do not affect the moments.  

The Glauber propagation  of the vector system $V$ produced by the diffractive DIS interaction on the front face and its inelastic interaction with the nucleons in the nuclear interior $V + N_b \to X$ occurs {\bf after} the lepton interacts with the struck quark.  
Because of the rescattering dynamics, the DDIS amplitude acquires a complex phase from Pomeron and Regge exchange;  thus final-state  rescattering corrections lead to  nontrivial ``dynamical" contributions to the measured PDFs; i.e., they involve physics aspects of the scattering process itself~\cite{Brodsky:2013oya}.  

Diffractive DIS is leading-twist and is the essential component of the two-step amplitude which causes shadowing and antishadowing of the nuclear PDF.  It is important to analyze whether the momentum and other sum rules derived from the OPE expansion in terms of local operators remain valid when these dynamical rescattering corrections to the nuclear PDF are included.   The OPE is derived assuming that the LF time separation between the virtual photons in the forward virtual Compton amplitude 
$\gamma^* A \to \gamma^* A$  scales as $1/Q^2$.
However, the propagation  of the vector system $V$ produced by the diffractive DIS interaction on the front face and its inelastic interaction with the nucleons in the nuclear interior $V + N_b \to X$ are characterized by a longer LF time  which scales as $ {1/W^2}$.  Thus the leading-twist multi-nucleon processes that produce shadowing and antishadowing in a nucleus are evidently not present in the $Q^2 \to \infty$ OPE analysis.

It should be emphasized  that shadowing in deep inelastic lepton scattering on a nucleus  involves the nucleons at or near the front surface -- the nucleons facing the incoming lepton beam. This  geometrical  bias is not built into the frame-independent nuclear LFWFs used to evaluate the matrix elements of local currents.  Thus the dynamical phenomena of leading-twist shadowing and antishadowing appear to invalidate the sum rules for nuclear PDFs.  The same complications occur in the leading-twist analysis of deeply virtual Compton scattering $\gamma^* A \to \gamma^* A$ on a nuclear target.
\section {Representing Inelastic Lepton-Proton Scattering as a Collision of Virtual Photon and Proton Structure Functions}

\begin{figure}
 \begin{center}
\includegraphics[height=9.5cm,width=15cm]{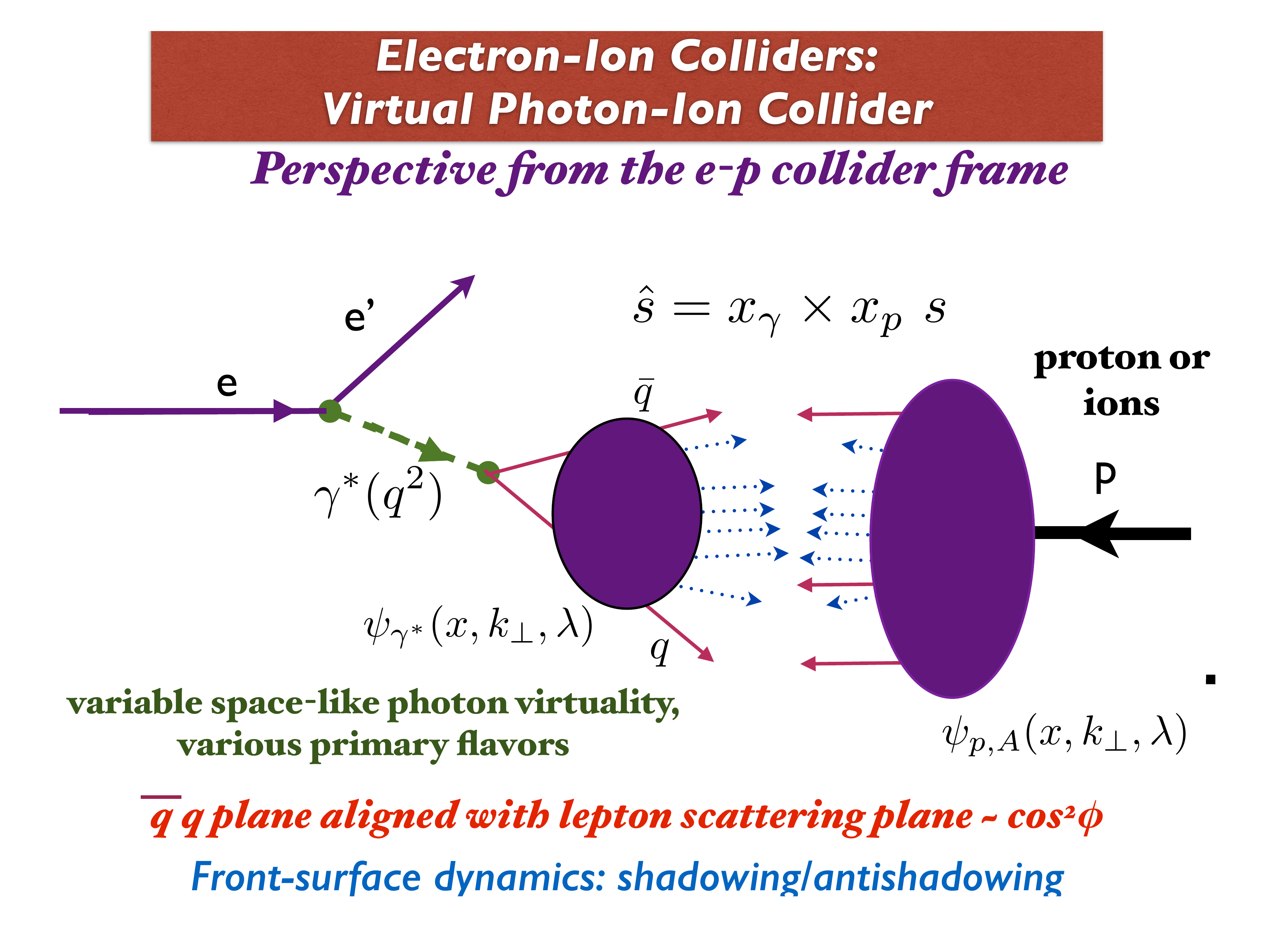}
\end{center}
\caption {Interactions at an electron-ion collider viewed as the collision of a photon structure function  (real or virtual) with a proton or nuclear structure function. }
\label{FigsJlabProc13.pdf}
\end{figure}

It is useful to think of electron-proton collisions at JLab in terms of the collision of the virtual photon structure function with the
proton structure function, as illustrated in fig.~\ref{FigsJlabProc13.pdf}.   For example, a  real or virtual photon emitted by the electron can 
readily couple to charm quark pairs $c \bar c$. The $c$ or $\bar c$  can then exchange a minimally off-shell gluon with a quark of the proton target, in analogy to Coulomb exchange in Bethe-Heitler $\tau$-pair production: $\gamma^* p \to \tau^+ \tau^- p$.   In the case of the LHeC, the virtual photon can first couple to top quark pairs which can exchange a minimally off-shell gluon and then emit a Higgs particle, etc.~\cite{Sarmiento-Alvarado:2014eha}.
This physics is obscured if one utilizes the conventional ``infinite momentum frame"  $(q^+=q^0+q^3=0)$ description,  where the dynamics of the collision  is attributed solely to the proton's parton distribution function.

The collision of structure functions illustrated in fig. \ref{FigsJlabProc13.pdf} can be used to understand the production of high multiplicity events in electron-proton collisions as due to the interactions of the gluonic flux tube connecting the $q$ and $\bar q$ of the virtual photon with the flux tube between the $q$ and $qq$ diquark of the proton. Maximum multiplicity occurs when the flux tubes are aligned at the same  azimuthal angle $\phi$.  As Bjorken, Goldhaber and I have discussed~\cite{Bjorken:2013boa}, one can understand the occurrence of the ``ridge" phenomena produced at RHIC in proton-proton collisions from this perspective: high multiplicity events caused by the activations of the colliding flux tubes will extend over all rapidities on both the trigger side and away side.  In the case of $e p \to e^\prime X$, the $ q \bar q$ plane  created by the virtual photon is aligned with the electron scattering plane. Thus the ridge of the hadronic final state produced in 
high-multiplicity $\gamma^* p$ collisions will tend to be aligned  with the electron scattering plane.  The properties of the $q \bar q$ flux tube will be controlled both by the flavor of the $q \bar q $ pair and the photon virtuality $Q^2$. Thus electron-proton collisions provide an important laboratory to study  the physics underlying ridge phenomena in QCD.

\section{Charm Production at Jab12}

\begin{figure}
 \begin{center}
\includegraphics[height=6cm,width=7.5cm]{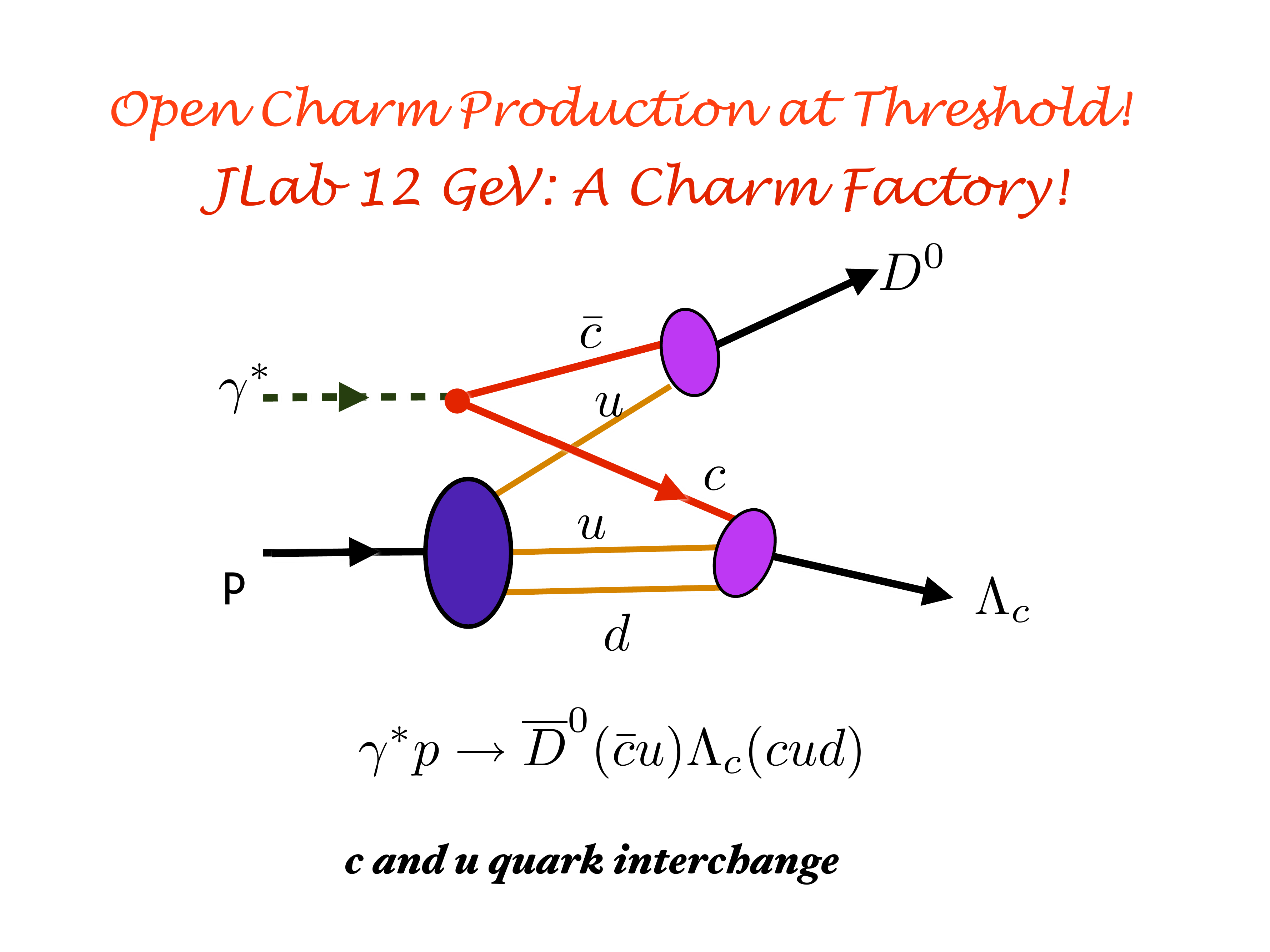}
\includegraphics[height=6cm,width=7.5cm]{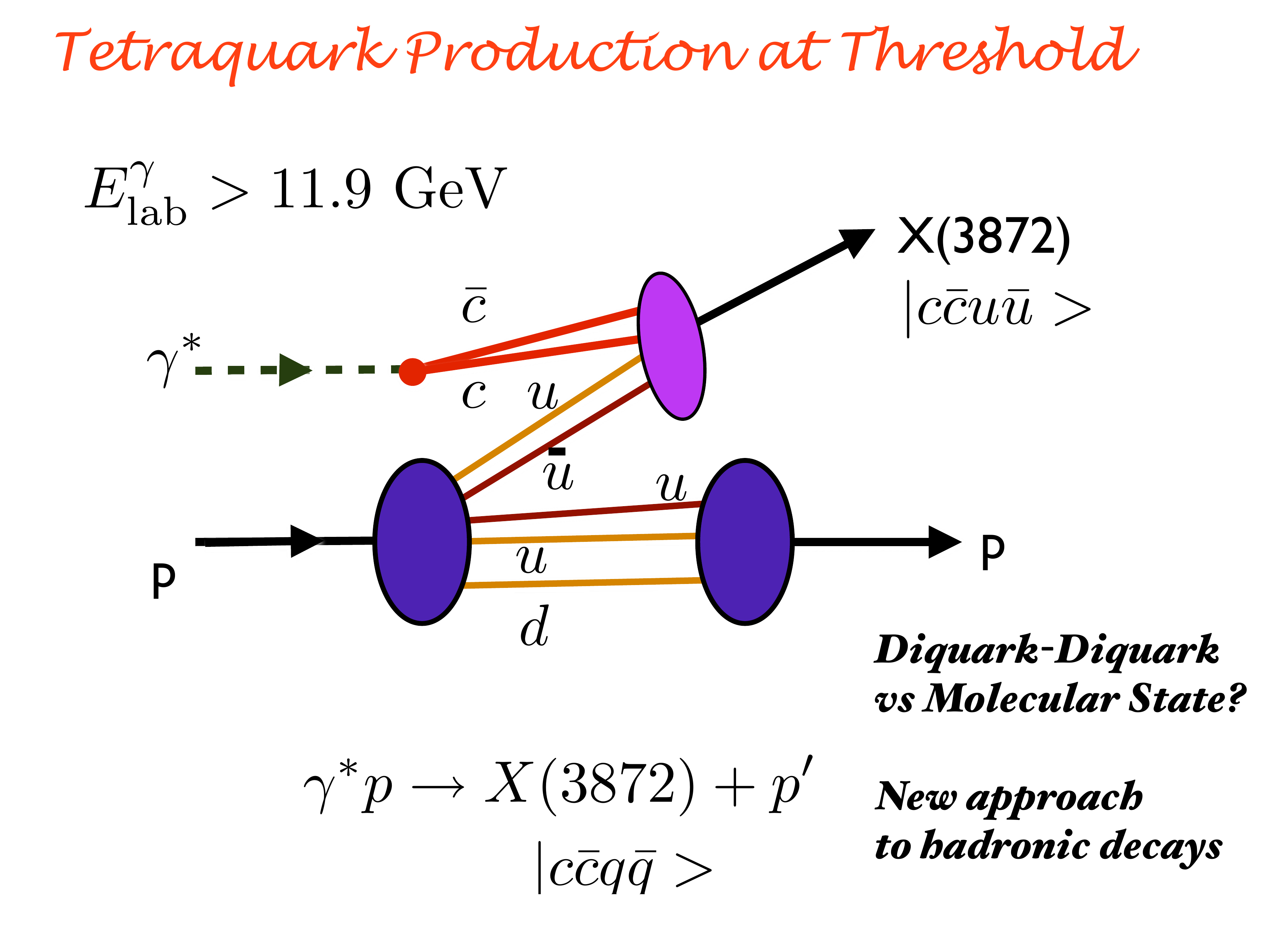}
\includegraphics[height=6cm,width=7.5cm]{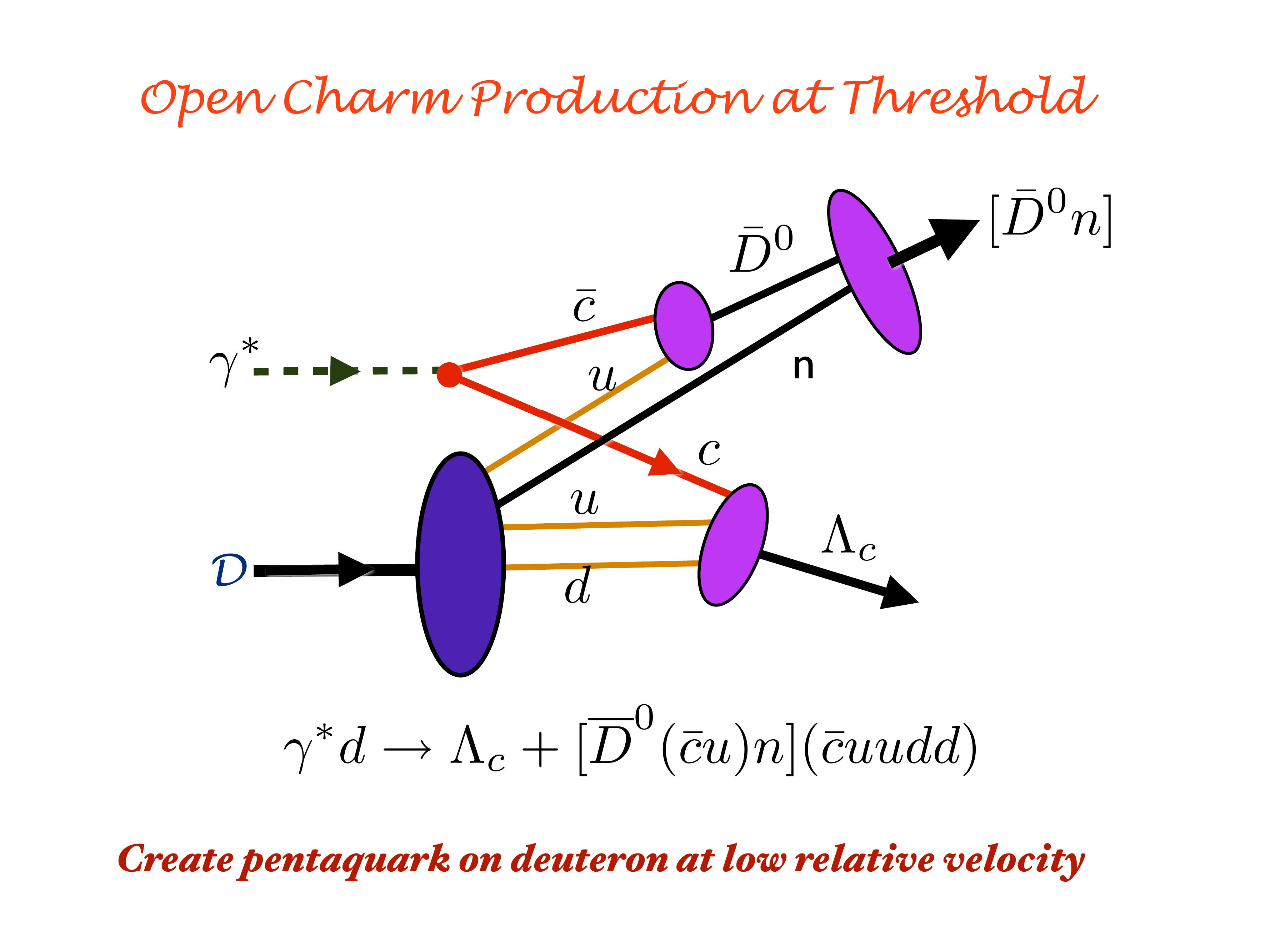}
\includegraphics[height=7cm,width=7.5cm]{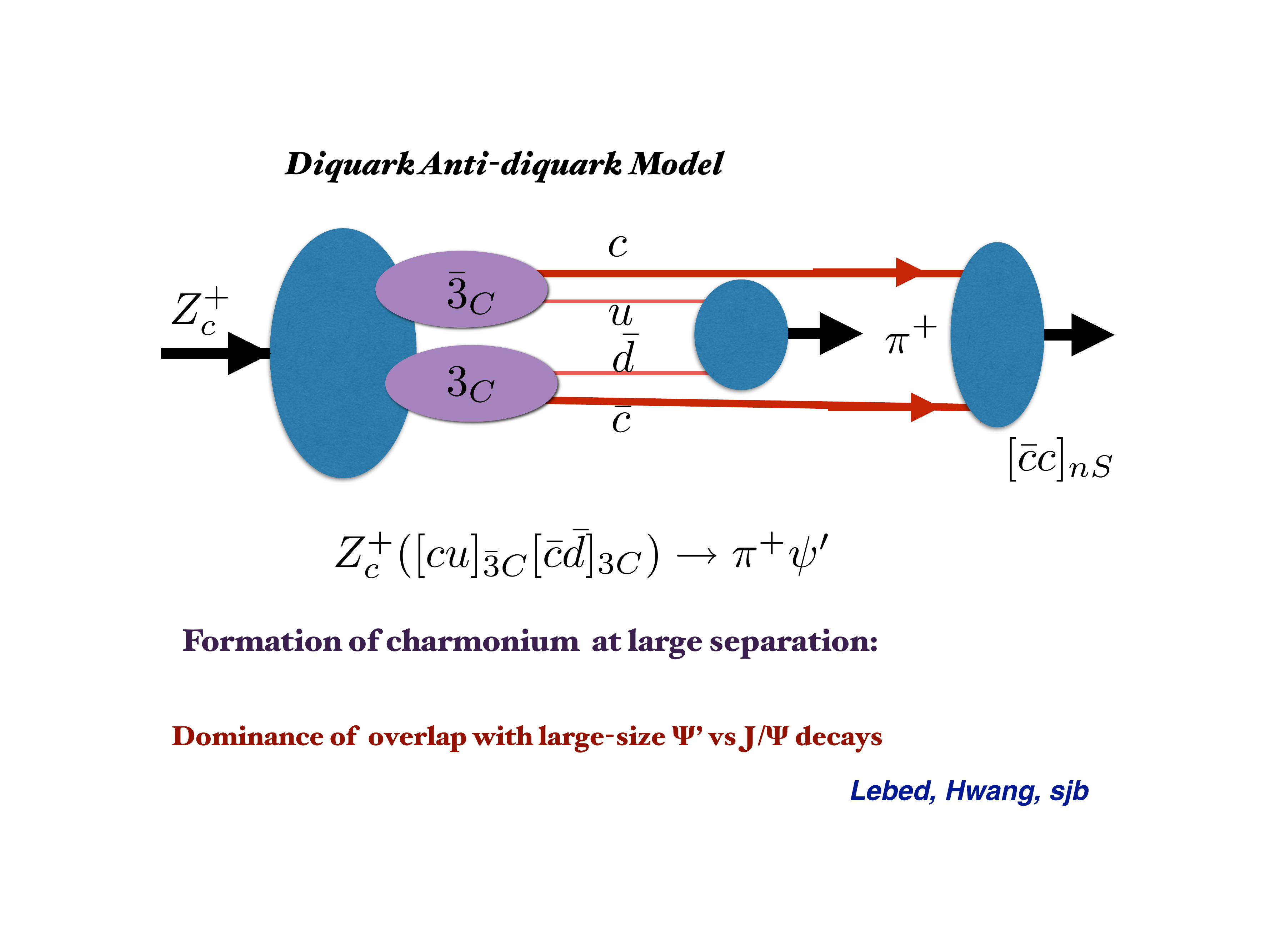}
\end{center}
\caption{ (A) Open charm can be electroproduced or photoproduced at JLab12.  The quark-interchange amplitude for a typical open charm exclusive channel $\gamma^* p \to \bar D \Lambda_c$ is illustrated.   
(B) Contribution to the exclusive electroproduction of a tetraquark ~\cite{Brodsky:2015wza}.
(C) Exclusive electroproduction of a $|\bar c  u udd > $  pentaquark on a deuteron target. 
(D) Model for the decay of the $Z^+_c(c u \bar c \bar d)$  tetraquark to a pion plus charmonium state if it is a bound state composed of a $[\bar c u]$ $3_C$  diquark and  a $ [ c \bar d]$  $3_C$  antidiquark.  
The $Z^+_c$  decay begins to occur when the $u$ and $\bar d$ overlap to form the $\pi^+$.  Since this occurs primarily when the diquark and antidiquark match the size of the pion, the $c$ and $\bar c$ will also be at relatively large separation. Thus the $c$ and  $\bar c$ are more likely to form  the larger size  $\psi^\prime $ quarkonium state than the more compact $J/\psi.$  See ref.~\cite{Brodsky:2015wza}. 
}
\label{FigsJlabProcFig7.pdf}
\end{figure}

The available  CM energy $s = (q+p)^2 = M^2 + q^2 + 2  M_p E_\gamma$ at JLab-12 is above the open charm threshold.   The $u$-quark-interchange amplitude for a typical open charm exclusive channel $\gamma^* p \to \bar D^0(\bar c u)  \Lambda_c(cdu)$ is illustrated  in fig. \ref{FigsJlabProcFig7.pdf}(A).  When one  is close to threshold,  
$W^2 =(q+p)^2 \simeq ( M_p + M_{J/\psi} )^2$,   diagrams where the $c \bar c$ pair is multiple-connected to the valence quarks of the proton become dominant. The multi-connected diagrams are the basis for intrinsic charm~\cite{Brodsky:1984nx,Brodsky:1980pb,Brodsky:2015uwa}, which dominates the charm structure function at large $x_{bj}$. It is also advantageous to utilize the Fermi motion of a nuclear target. 
The cross section for open charm is further enhanced near threshold because of the Coulombic gluonic interactions between the $c$ and  $\bar c$. 
The rate is enhanced at threshold by the Sudakov-Sakharov-Sommerfeld effect due to multi-gluon exchange at the soft scale  of order $v^2 s$, where $v$ is the relative velocity of the $c \bar c$~\cite{Brodsky:1995ds}.  This analysis of multi-scale processes uses BLM/PMC renormalization scale setting~\cite{Brodsky:1982gc,Mojaza:2012mf}. The electroproduction of the $J/\psi$ was observed with a substantial cross section at Cornell close to threshold~\cite{Gittelman:1975ix,Brodsky:2000zc}. 

Tetraquarks such as the $X^0(3872)|c \bar c u \bar u>$  can also be electroproduced near threshold in exclusive reactions 
such as $\gamma^* p \to X^0 p$~\cite{Brodsky:2015wza,Lebed:2015sxa},  as illustrated in fig. \ref{FigsJlabProcFig7.pdf} B.  Features of the production cross section can be used to analyze the composition of the tetraquark as a $3_C \bar 3_C$ diquark-antidiquark bound state~\cite{Maiani:2004vq,Brodsky:2015wza} or a meson-meson molecule~\cite{Karliner:2013gma}, in analogy to nuclear-bound quarkonium~\cite{Brodsky:1989jd, Luke:1992tm}.    
The strong attractive color binding of  the diquark-antidiquark model gives maximal binding. 
As Hwang, Lebed, and I have shown~\cite{Brodsky:2015wza}  the $3_C \bar 3_C$ diquark-antidiquark structure can account for the observed dominance of the $Z^+_c \to \psi^\prime \pi^+$ decay compared to  $Z^+_c \to  J/\psi \pi^+$ decay.  The mechanism is illustrated in fig. \ref{FigsJlabProcFig7.pdf}D. 
The electroproduction of a charmed $|\bar c  u udd > $ pentaquark on a a deuteron target $\gamma^* {\cal D} \to  [\bar D^0 n] \Lambda_c$  is illustrated in fig \ref{FigsJlabProcFig7.pdf}C. 
It is even possible to electroproduce an  ``octoquark" bound state~\cite{Bashkanov:2013cla} with eight quarks $|\bar c  c uud udd > $ on a deuteron target $e  {\cal D} \to  [\bar D^0 n] \Lambda_c e$.   The existence of the  $|\bar c  c uud uud > $ octoquark can account for the strong $A_{NN}$ transverse  spin correlation observed at the charm threshold in polarized $p^\updownarrow p^\updownarrow \to pp$ elastic scattering. See ref.~\cite{Brodsky:1987xw,Broadsky:2012rw} 

 \section{Color Confinement and Supersymmetry in Hadron Physics from LF Holography}

\begin{figure}
 \begin{center}
\includegraphics[height= 9cm,width=15cm]{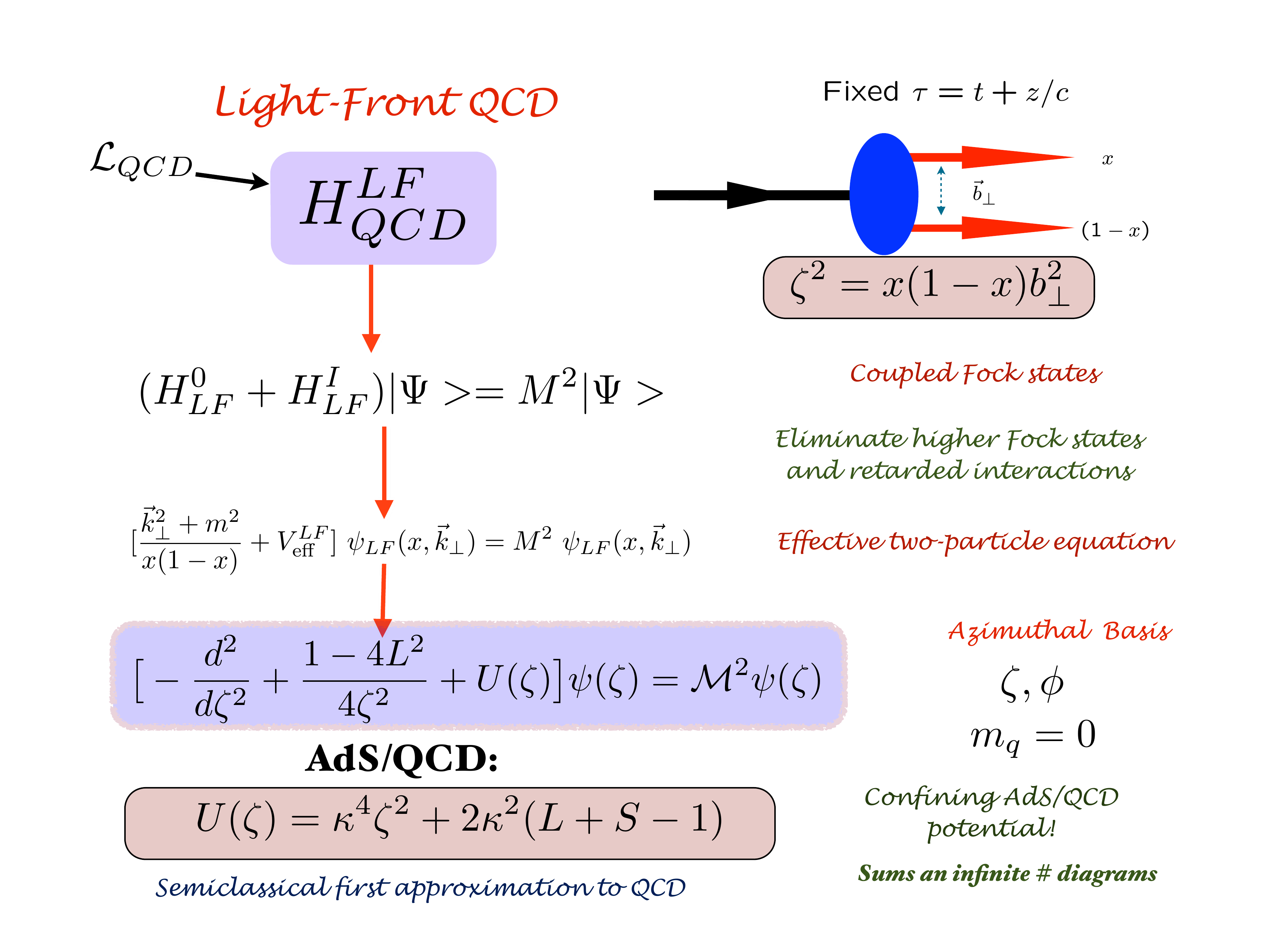}
\end{center}
\caption{Derivation of the Effective Light-Front Schr\"odinger Equation from QCD.  As in QED, one reduces the LF Heisenberg equation $H_{LF}|\Psi >   = M^2 |\Psi>$ 
to an effective two-body eigenvalue equation for $q \bar q$ mesons by systematically eliminating higher Fock states. One utilizes the LF radial variable $\zeta$, where $\zeta^2 = x(1-x)b^2_\perp$ is conjugate to the $q \bar q$ LF kinetic energy $k^2_\perp\over x(1-x)$ for $m_q=0$ to reduce the dynamics to a single variable bound state equation.  The confining potential $U(\zeta)$, including its spin-J dependence, is derived from the soft-wall AdS/QCD model with the dilaton  $e^{+\kappa^2 z^2 },$ where $z$ is the fifth coordinate of AdS$_5$ holographically dual  to $\zeta$. See ref.~\cite{Brodsky:2013ar}.   The light-front harmonic oscillator confinement potential $\kappa^4 \zeta^2 $ for is equivalent to a linear confining potential for heavy quarks in the instant form~\cite{Trawinski:2014msa}. }
\label{FigsJlabProcFig2.pdf}
\end{figure}

One of the most fundamental problems in quantum chromodynamics is to understand the origin of the mass scale which controls the range of color confinement and the hadronic spectrum.  For example, if one sets the Higgs couplings of quarks to zero, then no mass parameter appears in the QCD Lagrangian, and the theory is conformal at the classical level.   Nevertheless,  hadrons have a finite mass.  de Teramond, Dosch, and I~\cite{Brodsky:2013ar}
have shown that a mass gap and a fundamental color confinement scale can be derived from a conformally covariant action when one extends the formalism of de Alfaro, Fubini and Furlan to light-front Hamiltonian theory. Remarkably, the resulting light-front potential has a unique form of a harmonic oscillator $\kappa^4 \zeta^2$ in the 
light-front invariant impact variable $\zeta$ where $ \zeta^2Ê = b^2_\perp x(1-x)$. The result is  a single-variable frame-independent relativistic equation of motion for  $q \bar q$ bound states, a ``Light-Front Schr\"odinger Equation"~\cite{deTeramond:2008ht}, analogous to the nonrelativistic radial Schr\"odinger equation in quantum mechanics.  The  Light-Front Schr\"odinger Equation  incorporates color confinement and other essential spectroscopic and dynamical features of hadron physics, including a massless pion for zero quark mass and linear Regge trajectories with the same slope  in the radial quantum number $n$   and internal  orbital angular momentum $L$.   
The same light-front  equation for mesons of arbitrary spin $J$ can be derived~\cite{deTeramond:2013it}
from the holographic mapping of  the ``soft-wall model" modification of AdS$_5$ space with the specific dilaton profile $e^{+\kappa^2 z^2}$,  where one identifies the fifth dimension coordinate $z$ with the light-front coordinate $\zeta$.  The five-dimensional AdS$_5$ space provides a geometrical representation of the conformal group.
It is holographically dual to 3+1  spacetime using light-front time $\tau = t+ z/c$.  The derivation of the confining LF Schrodinger Equation is outlined in fig. \ref{FigsJlabProcFig2.pdf}.

\begin{figure}
 \begin{center}
\includegraphics[height=6cm,width=7.5cm]{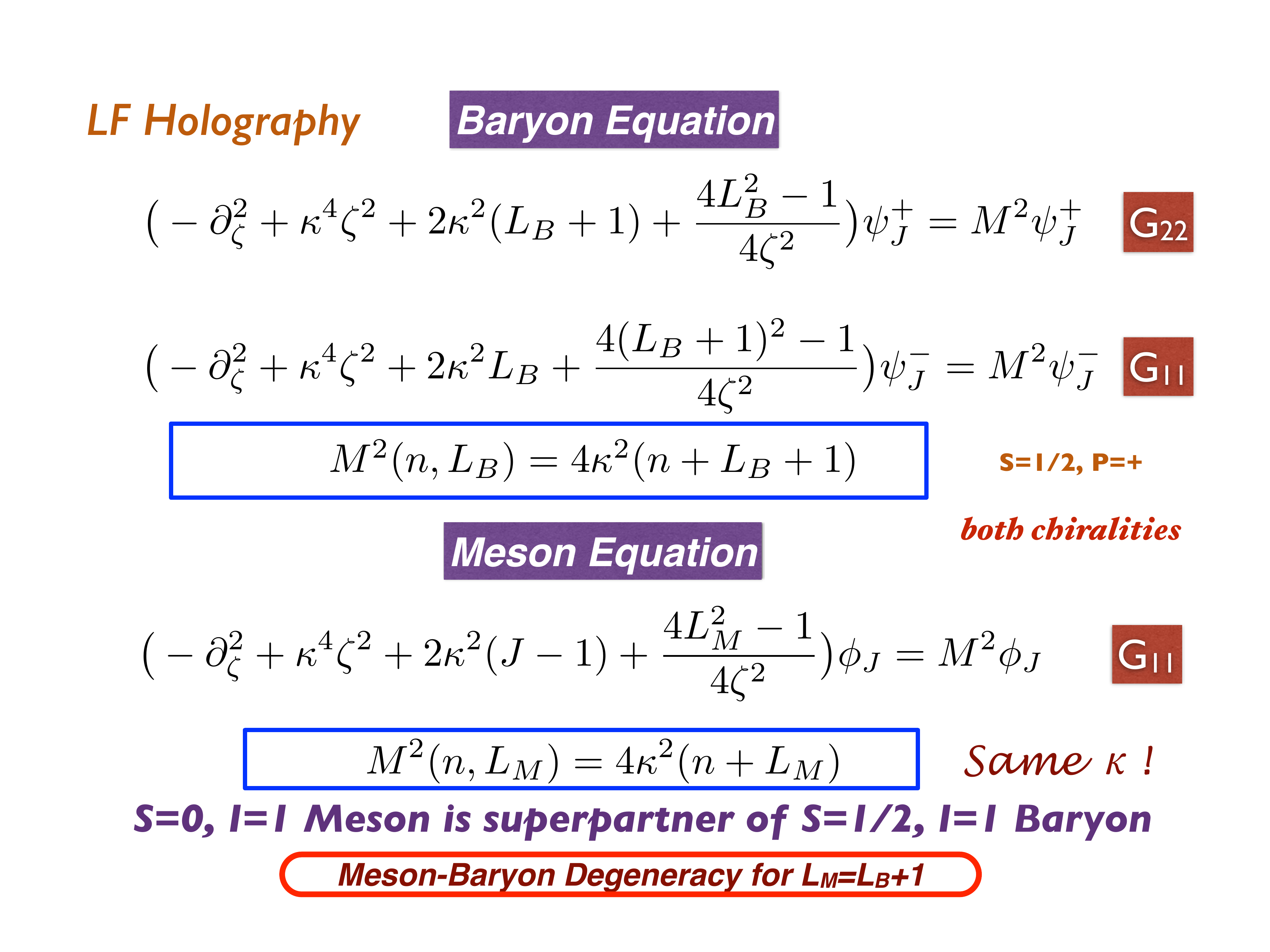}
\includegraphics[height=6cm,width=7.5cm]{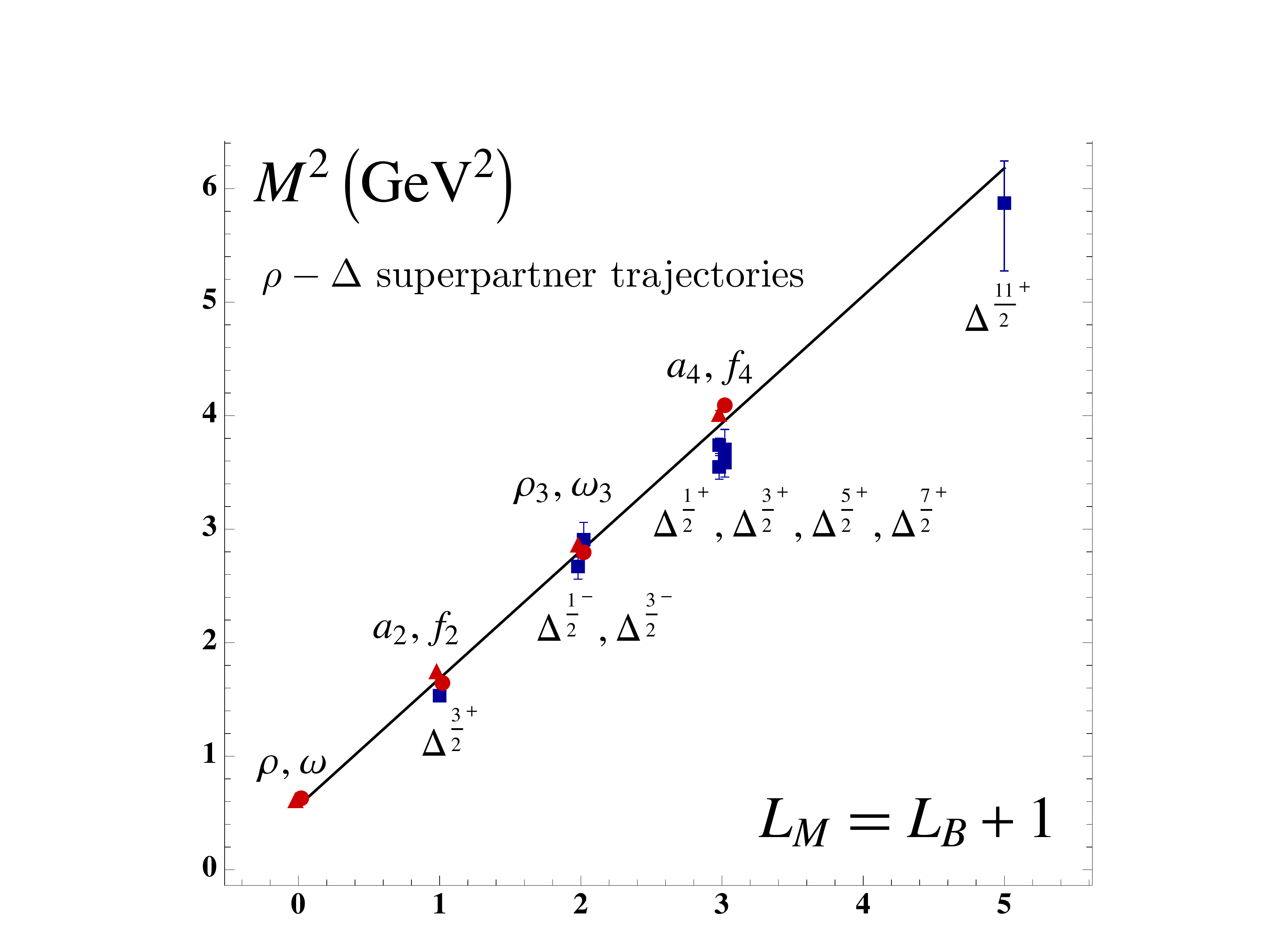}
\end{center}
\caption{(A). The LF Schr\"odinger equations for baryons and mesons for zero quark mass derived from the Pauli $2\times 2$ matrix representation of superconformal algebra.  
The $\psi^\pm$  are the baryon quark-diquark LFWFs where the quark spin $S^z_q=\pm 1/2$ is parallel or antiparallel to the baryon spin $J^z=\pm 1/2$.   The meson and baryon equations are identical if one identifies a meson with internal orbital angular momentum $L_M$ with its superpartner baryon with $L_B = L_M-1.$
See ref.~\cite{deTeramond:2014asa,Dosch:2015nwa,Dosch:2015bca}.
(B). Comparison of the $\rho/\omega$ meson Regge trajectory with the $J=3/2$ $\Delta$  baryon trajectory.   Superconformal algebra  predicts the degeneracy of the  meson and baryon trajectories if one identifies a meson with internal orbital angular momentum $L_M$ with its superpartner baryon with $L_M = L_B+1.$
See refs.~\cite{deTeramond:2014asa,Dosch:2015nwa}.}
\label{FigsJlabProcFig3.pdf}
\end{figure}

Thus  the combination of light-front dynamics, its holographic mapping to AdS$_5$ space, and the dAFF procedure provides new  insight into the physics underlying color confinement, the nonperturbative QCD coupling, and the QCD mass scale.  A comprehensive review is given in ref.~\cite{Brodsky:2014yha}.  The $q \bar q$ mesons and their valence LF wavefunctions are the eigensolutions of a frame-independent bound state equation, the Light-Front Schr\"odinger Equation.  The mesonic $q\bar q$ bound-state eigenvalues for massless quarks are $M^2(n, L, S) = 4\kappa^2(n+L +S/2)$.
The equation predicts that the pion eigenstate  $n=L=S=0$ is massless at zero quark mass, The  Regge spectra of the pseudoscalar $S=0$  and vector $S=1$  mesons  are 
predicted correctly, with equal slope in the principal quantum number $n$ and the internal orbital angular momentum.  The predicted nonperturbative pion distribution amplitude 
$\phi_\pi(x) \propto f_\pi \sqrt{x(1-x)}$ is  consistent with the Belle data for the photon-to-pion transition form factor~\cite{Brodsky:2011xx}. The prediction for the LFWF $\psi_\rho(x,k_\perp)$ of the  $\rho$ meson gives excellent 
predictions for the observed features of diffractive $\rho$ electroproduction $\gamma^* p \to \rho  p^\prime$~\cite{Forshaw:2012im}.
These results can be extended~\cite{deTeramond:2014asa,Dosch:2015nwa,Dosch:2015bca} to effective QCD light-front equations for both mesons and baryons by using the generalized supercharges of superconformal algebra~\cite{Fubini:1984hf}. 
The supercharges connect the baryon and meson spectra  and their Regge trajectories to each other in a remarkable manner: each meson has internal  angular momentum one unit higher than its superpartner baryon.  See  fig. \ref {FigsJlabProcFig3.pdf}(A).   Only one mass parameter $\kappa$ appears; it sets the confinement and the hadron mass scale in the  chiral limit, as well as  the length scale which underlies hadron structure.  ``Light-Front Holography"  not only predicts meson and baryon  spectroscopy  successfully, but also hadron dynamics:  light-front wavefunctions, vector meson electroproduction, distribution amplitudes, form factors, and valence structure functions.  
The LF Schr\"odinger Equations for baryons and mesons derived from superconformal algebra  are shown  in fig. \ref{FigsJlabProcFig3.pdf}.
The comparison between the meson and baryon masses of the $\rho/\omega$ Regge trajectory with the spin-$3/2$ $\Delta$ trajectory is shown in fig. \ref{FigsJlabProcFig3.pdf}(B).
Superconformal algebra  predicts the meson and baryon masses are identical if one identifies a meson with internal orbital angular momentum $L_M$ with its superpartner baryon with $L_B = L_M-1.$   Notice that the twist  $\tau = 2+ L_M = 3 + L_B$ of the interpolating operators for the meson and baryon superpartners are the same.   Superconformal algebra also predicts that the LFWFs of the superpartners are identical, and thus they have identical dynamics, such their elastic and transition form factors.   These features can be tested for spacelike  form factors at  JLab12.
\begin{figure}
\begin{center}
\includegraphics[height=6cm,width=7.5cm]{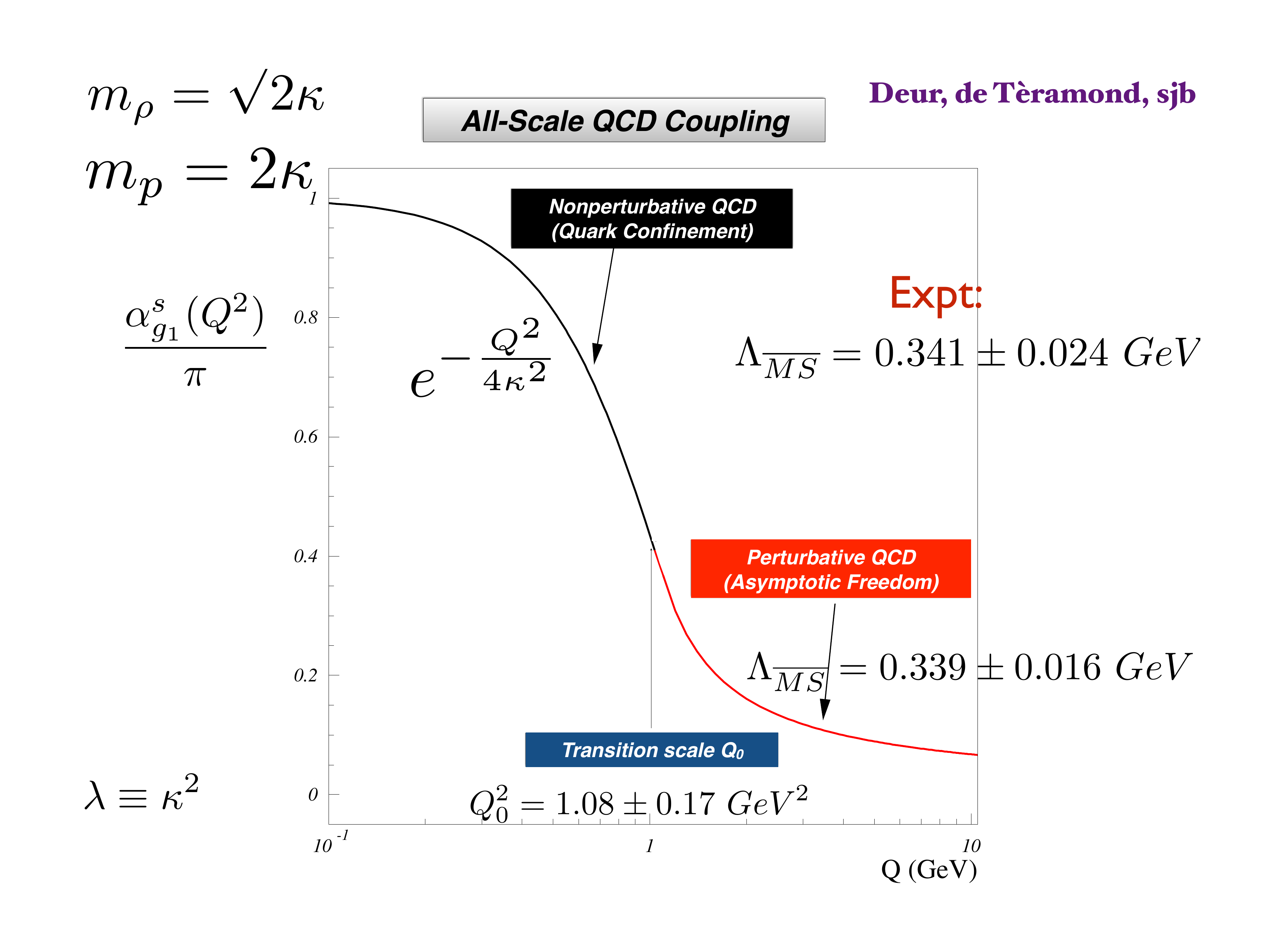}
\includegraphics[height=6cm,width=7.5cm]{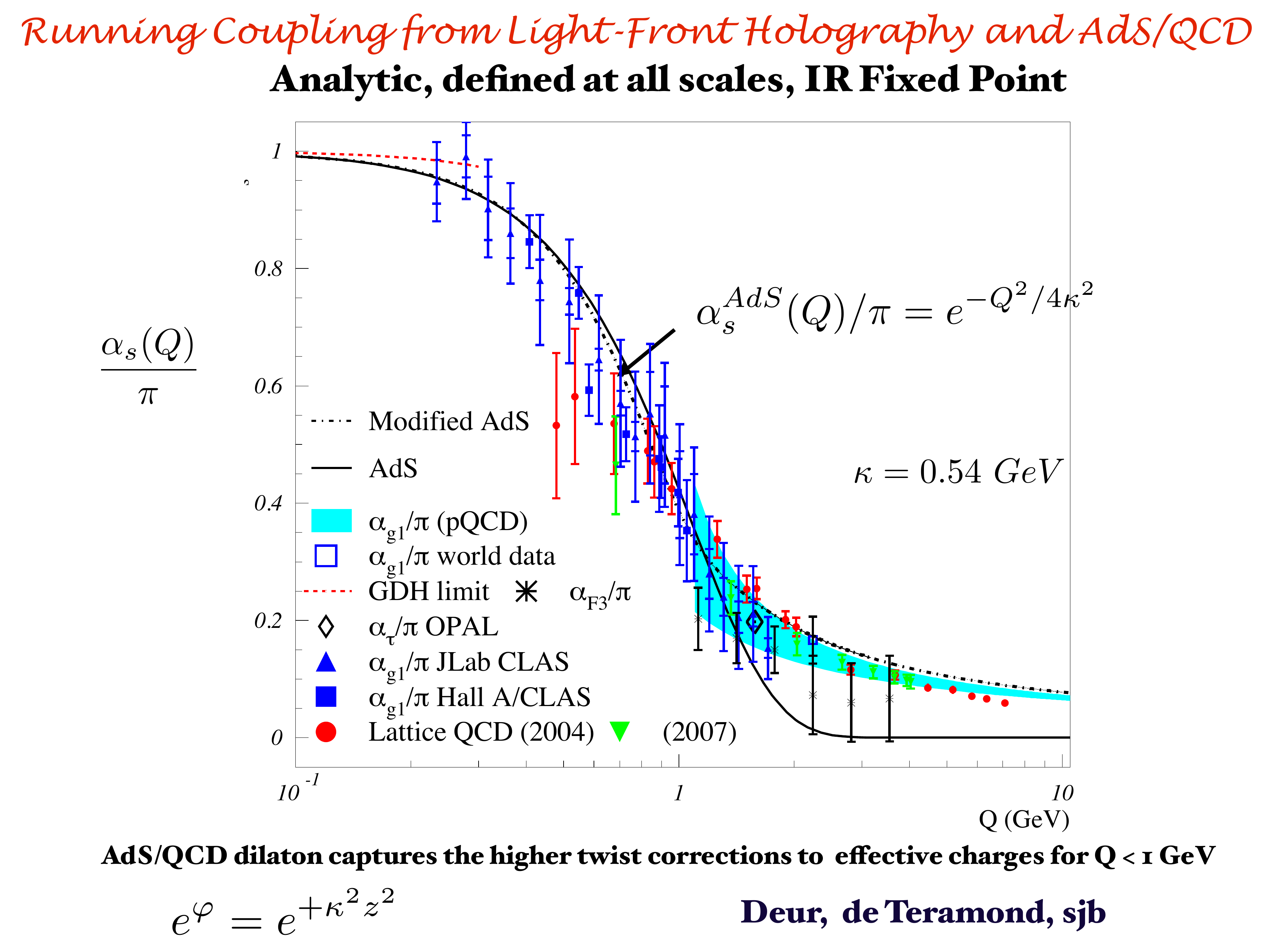}
\end{center}
\caption{
(A)  Prediction from LF Holography for the QCD Running Coupling $\alpha^s_{g_1}(Q^2)$.   The magnitude and derivative of the perturbative and nonperturbative coupling are matched at the scale $Q_0$.  This matching connects the perturbative scale 
$\Lambda_{\overline{MS}}$ to the nonpertubative scale $\kappa$ which underlies the hadron mass scale. 
(B) Comparison of the predicted nonpertubative coupling with measurements of the effective charge $\alpha^s_{g_1}(Q^2)$  
defined from the Bjorken sum rule.  
See ref.~\cite{Brodsky:2014jia}. 
}
\label{FigsJlabProcFig5.pdf}
\end{figure} 
\section {The QCD Coupling at all Scales} 
As Grunberg~\cite{Grunberg:1980ja} has emphasized, the QCD running coupling can be defined at all momentum scales from a perturbatively calculable observable, such as the coupling $\alpha^s_{g_1}(Q^2)$ which is defined from measurements of the Bjorken sum rule.   At high momentum transfer, such ``effective charges" satisfy asymptotic freedom, obey the usual pQCD renormalization group equations, and can be related to each other without scale ambiguity 
by commensurate scale relations~\cite{Brodsky:1994eh}.  
The dilaton  $e^{+\kappa^2 z^2}$ soft-wall modification of the AdS$_5$ metric, together with LF holography, predicts the functional behavior 
in the small $Q^2$ domain~\cite{Brodsky:2010ur}: 
${\alpha^s_{g_1}(Q^2) = 
\pi   e^{- Q^2 /4 \kappa^2 }}. $ 
Measurements of  $\alpha^s_{g_1}(Q^2)$ are remarkably consistent with the predicted Gaussian form. 
Deur, de Teramond, and I~\cite{Deur:2014qfa,Brodsky:2010ur,Brodsky:2014jia} have also shown how the parameter $\kappa$,  which   determines the mass scale of  hadrons in the chiral limit, can be connected to the  mass scale $\Lambda^s$  controlling the evolution of the perturbative QCD coupling. The connection can be done for any choice of renormalization scheme, such as the $\overline{MS}$ scheme,
as seen in  fig.~\ref{FigsJlabProcFig5.pdf}. 
The relation between scales is obtained by matching at a scale $Q^2_0$ the nonperturbative behavior of the effective QCD coupling, as determined from light-front holography, to the perturbative QCD coupling with  asymptotic freedom.
The result of this perturbative/nonperturbative matching is an effective QCD coupling  defined at all momenta.

\section{Acknowledgements}
Presented at the QCD Evolution 2015 Workshop, Jefferson Laboratory,  May 26-30, 2015, 
I thank James Bjorken, Alexandre Deur, Guy de Teramond,  Guenter Dosch, Susan Gardner, Fred Goldhaber,  Paul Hoyer, Dae Sung Hwang,  Rich Lebed, 
Simonetta Liuti, Cedric Lorce, Michael Peskin, and Ivan Schmidt
 for helpful conversations and suggestions.
This research was supported by the Department of Energy,  contract DE--AC02--76SF00515.  
SLAC-PUB-16381.


\begin{thebibliography}{99}


\bibitem{Smith:2009yk} 
  E.~S.~Smith,
 ``The 12-GeV JLab Upgrade Project,''
  Nucl.\ Phys.\ A {\bf 827}, 599C (2009)
  [arXiv:0901.3249 [nucl-ex]].


\bibitem{Brodsky:2015wza} 
  S.~J.~Brodsky and R.~F.~Lebed,
 ``QCD dynamics of tetraquark production,''
  Phys.\ Rev.\ D {\bf 91}, no. 11, 114025 (2015)
  [arXiv:1505.00803 [hep-ph]].


\bibitem{Brodsky:1987xw} 
  S.~J.~Brodsky and G.~F.~de Teramond,
 ``Spin Correlations, QCD Color Transparency and Heavy Quark Thresholds in Proton Proton Scattering,''
  Phys.\ Rev.\ Lett.\  {\bf 60}, 1924 (1988).


\bibitem{Bashkanov:2013cla} 
  M.~Bashkanov, S.~J.~Brodsky and H.~Clement,
 ``Novel Six-Quark Hidden-Color Dibaryon States in QCD,''
  Phys.\ Lett.\ B {\bf 727}, 438 (2013)
  [arXiv:1308.6404 [hep-ph]].


\bibitem{Brodsky:2012zzb} 
  S.~J.~Brodsky,
 ``Exotic Effects at the Charm Threshold and Other Novel Physics Topics at JLab-12 GeV,''
  SLAC-PUB-14985.


\bibitem{Brodsky:1989qz} 
  S.~J.~Brodsky and H.~J.~Lu,
 ``Shadowing and Antishadowing of Nuclear Structure Functions,''
  Phys.\ Rev.\ Lett.\  {\bf 64}, 1342 (1990).


\bibitem{Brodsky:2004qa} 
  S.~J.~Brodsky, I.~Schmidt and J.~J.~Yang,
 ``Nuclear antishadowing in neutrino deep inelastic scattering,''
  Phys.\ Rev.\ D {\bf 70}, 116003 (2004)
  [hep-ph/0409279].


\bibitem{Brodsky:2005ww} 
  S.~J.~Brodsky,
 ``Novel Nuclear Effects in QCD: Non-Universality of Nuclear Antishadowing and Hidden Color Phenomena,''
  AIP Conf.\ Proc.\  {\bf 792}, 279 (2005).
  
\bibitem{Deur:2014qfa} 
  A.~Deur, S.~J.~Brodsky and G.~F.~de Teramond,
  ``Connecting the Hadron Mass Scale to the Fundamental Mass Scale of Quantum Chromodynamics,''
  arXiv:1409.5488 [hep-ph].  To be published in Physics Letters B.


\bibitem{Brodsky:2010ur} 
  S.~J.~Brodsky, G.~F.~de Teramond and A.~Deur,
 ``Nonperturbative QCD Coupling and its $\beta$-function from Light-Front Holography,''
  Phys.\ Rev.\ D {\bf 81}, 096010 (2010)
  [arXiv:1002.3948 [hep-ph]].


\bibitem{Brodsky:2002ue} 
  S.~J.~Brodsky, P.~Hoyer, N.~Marchal, S.~Peigne and F.~Sannino,
 ``Structure functions are not parton probabilities,''
  Phys.\ Rev.\ D {\bf 65}, 114025 (2002)
  [hep-ph/0104291].


\bibitem{Brodsky:2002cx} 
  S.~J.~Brodsky, D.~S.~Hwang and I.~Schmidt,
 ``Final state interactions and single spin asymmetries in semiinclusive deep inelastic scattering,''
  Phys.\ Lett.\ B {\bf 530}, 99 (2002)
  [hep-ph/0201296].


\bibitem{Brodsky:1999mz} 
  S.~J.~Brodsky, J.~Rathsman and C.~Merino,
 ``Odderon-Pomeron interference,''
  Phys.\ Lett.\ B {\bf 461}, 114 (1999)
  [hep-ph/9904280].


\bibitem{Brodsky:2009bp} 
  S.~J.~Brodsky, F.~J.~Llanes-Estrada, J.~T.~Londergan and A.~P.~Szczepaniak,
 ``Reggeon Non-Factorizability and the J=0 Fixed Pole in DVCS,''
  arXiv:0906.5515 [hep-ph].
  
  
\bibitem{Brodsky:1973hm} 
  S.~J.~Brodsky, F.~E.~Close and J.~F.~Gunion,
  ``A Gauge - Invariant Scaling Model Of Current Interactions With Regge Behavior And Finite Fixed Pole Sum Rules,''
  Phys.\ Rev.\ D {\bf 8}, 3678 (1973).



\bibitem{Brodsky:2014yha} 
  S.~J.~Brodsky, G.~F.~de Teramond, H.~G.~Dosch and J.~Erlich,
 ``Light-Front Holographic QCD and Emerging Confinement,''
  Phys.\ Rept.\  {\bf 584}, 1 (2015)
  [arXiv:1407.8131 [hep-ph]].


\bibitem{Dosch:2015nwa} 
  H.~G.~Dosch, G.~F.~de Teramond and S.~J.~Brodsky,
 ``Superconformal Baryon-Meson Symmetry and Light-Front Holographic QCD,''
  Phys.\ Rev.\ D {\bf 91}, no. 8, 085016 (2015)
  [arXiv:1501.00959 [hep-th]].


\bibitem{Brodsky:1983vf} 
  S.~J.~Brodsky, C.~R.~Ji and G.~P.~Lepage,
 ``Quantum Chromodynamic Predictions for the Deuteron Form-Factor,''
  Phys.\ Rev.\ Lett.\  {\bf 51}, 83 (1983).


\bibitem{Brodsky:1988xz} 
  S.~J.~Brodsky and A.~H.~Mueller,
 ``Using Nuclei to Probe Hadronization in QCD,''
  Phys.\ Lett.\ B {\bf 206}, 685 (1988).


\bibitem{Brodsky:1984nx} 
  S.~J.~Brodsky, J.~C.~Collins, S.~D.~Ellis, J.~F.~Gunion and A.~H.~Mueller,
 ``Intrinsic Chevrolets At The Ssc,''
  DOE/ER/40048-21 P4, SLAC-PUB-15471.


\bibitem{Brodsky:1980pb} 
  S.~J.~Brodsky, P.~Hoyer, C.~Peterson and N.~Sakai,
 ``The Intrinsic Charm of the Proton,''
  Phys.\ Lett.\ B {\bf 93}, 451 (1980).


\bibitem{Brodsky:2015uwa} 
  S.~J.~Brodsky and S.~Gardner,
 ``Comment on "New Limits on Intrinsic Charm in the Nucleon from Global Analysis of Parton Distributions",''
  arXiv:1504.00969 [hep-ph].


\bibitem{Brodsky:2009gx} 
  S.~J.~Brodsky and R.~F.~Lebed,
 ``Production of the Smallest QED Atom: True Muonium (mu+ mu-),''
  Phys.\ Rev.\ Lett.\  {\bf 102}, 213401 (2009)
  [arXiv:0904.2225 [hep-ph]].


\bibitem{Banburski:2012tk} 
  A.~Banburski and P.~Schuster,
 ``The Production and Discovery of True Muonium in Fixed-Target Experiments,''
  Phys.\ Rev.\ D {\bf 86}, 093007 (2012)
  [arXiv:1206.3961 [hep-ph]].


\bibitem{AbelleiraFernandez:2012cc} 
  J.~L.~Abelleira Fernandez {\it et al.} [LHeC Study Group Collaboration],
 ``A Large Hadron Electron Collider at CERN: Report on the Physics and Design Concepts for Machine and Detector,''
  J.\ Phys.\ G {\bf 39}, 075001 (2012)
  [arXiv:1206.2913 [physics.acc-ph]].


\bibitem{Dirac:1949cp} 
  P.~A.~M.~Dirac,
 ``Forms of Relativistic Dynamics,''
  Rev.\ Mod.\ Phys.\  {\bf 21}, 392 (1949).


\bibitem{Brodsky:1997de} 
  S.~J.~Brodsky, H.~C.~Pauli and S.~S.~Pinsky,
 ``Quantum chromodynamics and other field theories on the light cone,''
  Phys.\ Rept.\  {\bf 301}, 299 (1998)
  [hep-ph/9705477].


\bibitem{Schienbein:2007fs} 
  I.~Schienbein, J.~Y.~Yu, C.~Keppel, J.~G.~Morfin, F.~Olness and J.~F.~Owens,
 ``Nuclear parton distribution functions from neutrino deep inelastic scattering,''
  Phys.\ Rev.\ D {\bf 77}, 054013 (2008)
  [arXiv:0710.4897 [hep-ph]].


\bibitem{deTeramond:2008ht} 
  G.~F.~de Teramond and S.~J.~Brodsky,
 ``Light-Front Holography: A First Approximation to QCD,''
  Phys.\ Rev.\ Lett.\  {\bf 102}, 081601 (2009)
  [arXiv:0809.4899 [hep-ph]].
  
\bibitem{deTeramond:2013it}
G.~F.~de Teramond, H.~G.~Dosch and S.~J.~Brodsky,
``Kinematical and Dynamical Aspects of Higher-Spin Bound-State Equations in Holographic QCD,''  Phys.\ Rev.\ D {\bf 87},  075005 (2013)
[arXiv:1301.1651 [hep-ph]].

  
\bibitem{Trawinski:2014msa} 
  A.~P.~Trawinski, S.~D.~Glazek, S.~J.~Brodsky, G.~F.~de Teramond and H.~G.~Dosch,
``Effective confining potentials for QCD,''  Phys.\ Rev.\ D {\bf 90}, no. 7, 074017 (2014) [arXiv:1403.5651 [hep-ph]].


\bibitem{Fubini:1984hf} 
  S.~Fubini and E.~Rabinovici,
 ``Superconformal Quantum Mechanics,''
  Nucl.\ Phys.\ B {\bf 245}, 17 (1984).


\bibitem{Nikolaev:1975vy} 
  N.~N.~Nikolaev and V.~I.~Zakharov,
 ``Parton Model and Deep Inelastic Scattering on Nuclei,''
  Phys.\ Lett.\ B {\bf 55}, 397 (1975).


\bibitem{Glauber:1955qq} 
  R.~J.~Glauber,
 ``Cross-sections in deuterium at high-energies,''
  Phys.\ Rev.\  {\bf 100}, 242 (1955).


\bibitem{Stodolsky:1966am} 
  L.~Stodolsky,
 ``Hadron-like behavior of gamma, neutrino nuclear cross-sections,''
  Phys.\ Rev.\ Lett.\  {\bf 18}, 135 (1967).


\bibitem{Brodsky:1969iz} 
  S.~J.~Brodsky and J.~Pumplin,
 ``Photon-Nucleus Total Cross-Sections,''
  Phys.\ Rev.\  {\bf 182}, 1794 (1969).


\bibitem{Kuti:1971ph} 
  J.~Kuti and V.~F.~Weisskopf,
 ``Inelastic lepton - nucleon scattering and lepton pair production in the relativistic quark parton model,''
  Phys.\ Rev.\ D {\bf 4}, 3418 (1971).
  
\bibitem{Brodsky:1980zm} 
  S.~J.~Brodsky and S.~D.~Drell,
  ``The Anomalous Magnetic Moment and Limits on Fermion Substructure,''
  Phys.\ Rev.\ D {\bf 22}, 2236 (1980).
  


\bibitem{Liuti:2013cna} 
  S.~Liuti, A.~Rajan, A.~Courtoy, G.~R.~Goldstein and J.~O.~Gonzalez Hernandez,
 ``Partonic Picture of GTMDs,''
  Int.\ J.\ Mod.\ Phys.\ Conf.\ Ser.\  {\bf 25}, 1460009 (2014)
  [arXiv:1309.7029 [hep-ph]].
  

\bibitem{Mondal:2015uha} 
  C.~Mondal and D.~Chakrabarti,
  ``Generalized parton distributions and transverse densities in a light-front quarkÐdiquark model for the nucleons,''
  Eur.\ Phys.\ J.\ C {\bf 75}, no. 6, 261 (2015)
  [arXiv:1501.05489 [hep-ph]].
  
  
\bibitem{Lorce:2011dv} 
  C.~Lorce, B.~Pasquini and M.~Vanderhaeghen,
  ``Unified framework for generalized and transverse-momentum dependent parton distributions within a 3Q light-cone picture of the nucleon,''
  JHEP {\bf 1105}, 041 (2011)
  [arXiv:1102.4704 [hep-ph]].
  
\bibitem{Brodsky:2010vs} 
  S.~J.~Brodsky, B.~Pasquini, B.~W.~Xiao and F.~Yuan,
  ``Phases of Augmented Hadronic Light-Front Wave Functions,''
  Phys.\ Lett.\ B {\bf 687}, 327 (2010)
  [arXiv:1001.1163 [hep-ph]].



\bibitem{Brodsky:2008xe} 
  S.~J.~Brodsky,
 ``Dynamic versus Static Structure Functions and Novel Diffractive Effects in QCD,''
  AIP Conf.\ Proc.\  {\bf 1105}, 315 (2009)
  [arXiv:0811.0875 [hep-ph]].


\bibitem{Brodsky:2009dv} 
  S.~J.~Brodsky,
 ``Dynamic versus Static Hadronic Structure Functions,''
  Nucl.\ Phys.\ A {\bf 827}, 327C (2009)
  [arXiv:0901.0781 [hep-ph]].


\bibitem{Brodsky:2013oya} 
  S.~J.~Brodsky, D.~S.~Hwang, Y.~V.~Kovchegov, I.~Schmidt and M.~D.~Sievert,
 ``Single-Spin Asymmetries in Semi-inclusive Deep Inelastic Scattering and Drell-Yan Processes,''
  Phys.\ Rev.\ D {\bf 88}, no. 1, 014032 (2013)
  [arXiv:1304.5237 [hep-ph]].




\bibitem{Sarmiento-Alvarado:2014eha} 
  I.~A.~Sarmiento-Alvarado, A.~O.~Bouzas and F.~Larios,
 ``Analysis of top-quark charged-current coupling at the LHeC,''
  J.\ Phys.\ G {\bf 42}, no. 8, 085001 (2015)
  [arXiv:1412.6679 [hep-ph]].


\bibitem{Bjorken:2013boa} 
  J.~D.~Bjorken, S.~J.~Brodsky and A.~Scharff Goldhaber,
 ``Possible multiparticle ridge-like correlations in very high multiplicity proton-proton collisions,''
  Phys.\ Lett.\ B {\bf 726}, 344 (2013)
  [arXiv:1308.1435 [hep-ph]].


\bibitem{Brodsky:1995ds} 
  S.~J.~Brodsky, A.~H.~Hoang, J.~H.~Kuhn and T.~Teubner,
 ``Angular distributions of massive quarks and leptons close to threshold,''
  Phys.\ Lett.\ B {\bf 359}, 355 (1995)
  [hep-ph/9508274].


\bibitem{Brodsky:1982gc} 
  S.~J.~Brodsky, G.~P.~Lepage and P.~B.~Mackenzie,
 ``On the Elimination of Scale Ambiguities in Perturbative Quantum Chromodynamics,''
  Phys.\ Rev.\ D {\bf 28}, 228 (1983).


\bibitem{Mojaza:2012mf} 
  M.~Mojaza, S.~J.~Brodsky and X.~G.~Wu,
 ``Systematic All-Orders Method to Eliminate Renormalization-Scale and Scheme Ambiguities in Perturbative QCD,''
  Phys.\ Rev.\ Lett.\  {\bf 110}, 192001 (2013)
  [arXiv:1212.0049 [hep-ph]].


\bibitem{Gittelman:1975ix} 
  B.~Gittelman, K.~M.~Hanson, D.~Larson, E.~Loh, A.~Silverman and G.~Theodosiou,
 ``Photoproduction of the psi (3100) Meson at 11-GeV,''
  Phys.\ Rev.\ Lett.\  {\bf 35}, 1616 (1975).


\bibitem{Brodsky:2000zc} 
  S.~J.~Brodsky, E.~Chudakov, P.~Hoyer and J.~M.~Laget,
 ``Photoproduction of charm near threshold,''
  Phys.\ Lett.\ B {\bf 498}, 23 (2001)
  [hep-ph/0010343].


\bibitem{Lebed:2015sxa} 
  R.~F.~Lebed,
 ``A New Dynamical Picture for the Production and Decay of the $XY \! Z$ Mesons,''
  arXiv:1508.03320 [hep-ph].


\bibitem{Maiani:2004vq} 
  L.~Maiani, F.~Piccinini, A.~D.~Polosa and V.~Riquer,
 ``Diquark-antidiquarks with hidden or open charm and the nature of X(3872),''
  Phys.\ Rev.\ D {\bf 71}, 014028 (2005)
  [hep-ph/0412098].


\bibitem{Karliner:2013gma} 
  M.~Karliner, H.~J.~Lipkin and N.~A.~T\"ornqvist,
 ``New hadrons with heavy quarks,''
  Acta Phys.\ Polon.\ Supp.\  {\bf 6}, 181 (2013).


\bibitem{Brodsky:1989jd} 
  S.~J.~Brodsky, I.~A.~Schmidt and G.~F.~de Teramond,
 ``Nuclear Bound Quarkonium,''
  Phys.\ Rev.\ Lett.\  {\bf 64}, 1011 (1990).


\bibitem{Luke:1992tm} 
  M.~E.~Luke, A.~V.~Manohar and M.~J.~Savage,
 ``A QCD Calculation of the interaction of quarkonium with nuclei,''
  Phys.\ Lett.\ B {\bf 288}, 355 (1992)
  [hep-ph/9204219].


\bibitem{Broadsky:2012rw} 
  S.~Brodsky, G.~de Teramond and M.~Karliner,
 ``Puzzles in Hadronic Physics and Novel Quantum Chromodynamics Phenomenology,''
  Ann.\ Rev.\ Nucl.\ Part.\ Sci.\  {\bf 62}, 1 (2012)
  [arXiv:1302.5684 [hep-ph]].


\bibitem{Brodsky:2013ar} 
  S.~J.~Brodsky, G.~F.~De Teramond and H.~G.~Dosch,
 ``Threefold Complementary Approach to Holographic QCD,''
  Phys.\ Lett.\ B {\bf 729}, 3 (2014)
  [arXiv:1302.4105 [hep-th]].



\bibitem{deAlfaro:1976je} 
  V.~de Alfaro, S.~Fubini and G.~Furlan,
  ``Conformal Invariance in Quantum Mechanics,''
  Nuovo Cim.\ A {\bf 34}, 569 (1976).

\bibitem{Brodsky:2011xx} 
  S.~J.~Brodsky, F.~G.~Cao and G.~F.~de Teramond,
 ``Meson Transition Form Factors in Light-Front Holographic QCD,''
  Phys.\ Rev.\ D {\bf 84}, 075012 (2011)
  [arXiv:1105.3999 [hep-ph]].


\bibitem{Forshaw:2012im} 
  J.~R.~Forshaw and R.~Sandapen,
 ``An AdS/QCD holographic wavefunction for the rho meson and diffractive rho meson electroproduction,''
  Phys.\ Rev.\ Lett.\  {\bf 109}, 081601 (2012)
  [arXiv:1203.6088 [hep-ph]].


\bibitem{deTeramond:2014asa} 
  G.~F.~de Teramond, H.~G.~Dosch and S.~J.~Brodsky,
 ``Baryon Spectrum from Superconformal Quantum Mechanics and its Light-Front Holographic Embedding,''
  Phys.\ Rev.\ D {\bf 91}, no. 4, 045040 (2015)
  [arXiv:1411.5243 [hep-ph]].


\bibitem{Dosch:2015bca} 
  H.~G.~Dosch, G.~F.~de Teramond and S.~J.~Brodsky,
 ``Supersymmetry Across the Light and Heavy-Light Hadronic Spectrum,''
  arXiv:1504.05112 [hep-ph].


\bibitem{Grunberg:1980ja} 
  G.~Grunberg,
 ``Renormalization Group Improved Perturbative QCD,''
  Phys.\ Lett.\ B {\bf 95}, 70 (1980)
  [Phys.\ Lett.\ B {\bf 110}, 501 (1982)].


\bibitem{Brodsky:1994eh} 
  S.~J.~Brodsky and H.~J.~Lu,
 ``Commensurate scale relations in quantum chromodynamics,''
  Phys.\ Rev.\ D {\bf 51}, 3652 (1995)
  [hep-ph/9405218].


\bibitem{Brodsky:2014jia} 
  S.~J.~Brodsky, G.~F.~de Teramond, A.~Deur and H.~G.~Dosch,
 ``The Light-Front Schr\"odinger Equation and Determination of the Perturbative QCD Scale from Color Confinement,''
  arXiv:1410.0425 [hep-ph].




 




\end{thebibliography}
\end{document}